# Foundation-Model Earth Representations Enable Regional-Scale Forest Aboveground Biomass Monitoring Across the Northeastern United States

Shashika Lamahewage[1,*] and Chandi Witharana[1,2]

*[1] Department of Natural Resources and the Environment, Eversource Energy Center[2], University of Connecticut, Storrs. CT 06269, USA (shashika_himandi.lamahewa@uconn.edu[*]; chandi.witharana@uconn.edu)*

**Abstract**

Forest aboveground biomass (AGB) is a critical indicator of ecosystem productivity and terrestrial carbon storage, yet regional carbon monitoring remains constrained by the sparse spatial and temporal availability of field inventories and airborne structural measurements. Recent Earth observation foundation models provide globally consistent geospatial representations derived from diverse multimodal datasets, offering a potential pathway toward scalable biomass monitoring. Here, we evaluate Google Satellite Embeddings (GSE), generated by the AlphaEarth Foundation Model, for regional-scale AGB estimation across diverse temperate forest ecosystems in the northeastern United States. We integrated annual GSE observations, airborne LiDAR, and continuous forest inventory measurements from the Northeastern Forest Inventory Network (NEFIN) within a machine-learning framework. Combined LiDAR-GSE models achieved an $R^2$ of 0.79 for AGB estimation. Capitalizing on annual GSE observations expanded the training dataset by more than tenfold through temporal growth adjustment, increasing predictive performance to $R^2 = 0.82$ while reducing model bias by over 70%. Spatial autocorrelation analyses showed that integrating foundation-model representations and structural predictors substantially reduced residual spatial dependence. Monte Carlo simulations demonstrated that hyperparameter optimization reduced model-performance variability by 27.9%. Our findings demonstrate that foundation-model Earth representations capture ecologically meaningful information relevant to forest biomass and provide a scalable framework for annual carbon monitoring in regions with incomplete airborne LiDAR coverage. Our fundings establish a pathway toward next-generation forest carbon assessment based on globally available foundation-model Earth observations.

## Introduction

Forest ecosystems serve as a major carbon sink by sequestering atmospheric $CO_2$ in different ecosystem components for millennia. Forests store nearly 861 Gt of carbon, equivalent to almost a century of fossil fuel emissions [1]. About 42% of global carbon is stored in live vegetation as aboveground and belowground biomass [2]. Globally, temperate forests store the highest ecosystem aboveground carbon per acre accounting approximately for 34% [3]. Northeastern US (NEUSA) forests span over 70% of its landscape and serve as a significant carbon sink of the region. However, between 1990 and 2010, 31 ha of them were being converted to non-forest use every day [4]. Traditional field-based AGB estimation alone cannot keep pace with these accelerating environmental changes and rising data demands. Therefore, integrating traditional inventory data with state-of-the-art AI and remote sensing is essential to capture dynamics of carbon sequestration across space and time [5].

From the early 2000s researchers leaned towards remote sensing (RS) to obtain vegetation data due to fast, efficient and consistent coverage of Earth surface [6,7]. Often, RS data is calibrated by field based AGB estimates which use allometric equations for large area estimations [8,9]. LiDAR, synthetic aperture radar (SAR), satellite images (Landsat and Sentinel), and aerial imagery are among the most frequent data sources used in vegetation remote sensing [6, 10, 11]. As instances, Boudreau et al. [12] and Sheridan et al. [7] achieved $R^2$ of 0.65 and 0.87 respectively with discrete-return LiDAR point cloud data. Where, Puliti et al [13] combined Sentinel-2 and Landsat data to achieve $R^2$ values ranging from 0.24-0.64. However, each RS data source presents inherent limitations of its own. LiDAR is rich in 3-dimentional structural information yet limited in temporal and spatial distribution. Whereas RS imagery has limited sensitivity to vertical structure. Conversely, using multimodal RS inputs instead of single data source yields more accurate AGB estimates [5]. For instance, Xu et al. [14] used both optical and multi-temporal SAR datasets yielding $R^2$ of 0.69. In our recent work, Lamahewage et al [5] increased the $R^2$ from 0.23 to 0.41 by integrating LiDAR data with Sentinel-2 and National Agriculture Imagery Program (NAIP) data. Nandy et al. [15] observed an $R^2$ of 0.83 for the moist deciduous forests while combining ICESat-2 and Sentinel-1. However, inconsistencies in spatial and temporal resolution among traditional RS sources have prolonged the reproducibility of AGB estimations at regional and global scales in forest remote sensing.

In 2025, a remarkable breakthrough in RS applications redefined how multimodal Earth observations can be combined and analyzed. Google DeepMind's AlphaEarth Foundation (AEF) introduced 64 learned geospatial embeddings, expanding the horizons of multimodal data use in Earth science research [16, 17]. The Google Satellite Embedding (GSE) dataset provides a global analysis ready collection of learned geospatial embeddings, where each 10 m pixel is represented by a 64-dimensional feature vector [18]. GSE data are trained using diverse multimodal data sources, including optical (Sentinel-2, Landsat-8/9) images, thermal imagery, SAR (Sentinal-1), spaceborne LiDAR (GEDI), elevation, climate, gravity fields, land cover, and text-based information across varying spatial and temporal resolutions [16]. GSE dataset shows promising improvements in different classification, regression and time series tasks [19]. Despite these advances, GSE dataset remains largely uncharted in ecological research, particularly for AGB and forest carbon context [20]. Our study presents the first systematic framework to understand the utility of airborne LiDAR and the AlphaEarth GSE datasets for forests in the northeastern United States. Also, we aim to address the long-standing lack of consistent annual data at 10 m resolution in regional AGB estimations. To the best of our knowledge, our study marks the first step toward generating annual AGB estimates for the northeastern region covering more than 280 000 $km^2$ of the land at 15m spatial granularity.

Most RS datasets have been calibrated using field observations, such as the US Forest Service Forest Inventory and Analysis (FIA) data. Unfortunately, FIA true plot locations are no longer being shared with researchers, limiting their utility for spatial modeling and validation [5]. Therefore, most contemporary studies have opted to use alternative in-situ field data for ground truthing and model training [15]. Here, we introduce the Northeastern Forest Inventory Network's (NEFIN) continuous forest inventory (CFI) data as a substitute dataset for the regional AGB modeling [21]. Based on current knowledge, our study is the first to utilize NEFIN-CFI data in conjunction with airborne LiDAR and GSE within a regional AGB modelling framework.

Over the years, vast majority of RS based AGB modeling has rapidly evolved from traditional parametric models to nonparametric, machine learning (ML) approaches [22]. Regardless of the dataset, most publications are based on tree-based algorithms such as the random forest (RF), gradient boosting, and decision tree (CART) [21, 23]. For example, Hudak et al. [6], developed a carbon monitoring system utilizing 3805 plots and RF models across the northwestern US ($R^2$ = 0.80). Tang et al. [24] produced a forest carbon map with $R^2$ of 0.38 utilizing 1986 FIA subplot locations based on quantile RF approach. Srivastava and Pant [23] compared ML algorithms including Extremely Randomized Trees (ExtraTrees) and K Neighbors Regressor (KN). Additionally, Ma et al. [26] produced a 90 m resolution AGB map employing the Ecosystem Demography model. Blackard et al. [27], developed a 250 m resolution map by using tree-based algorithms, and another AGB map was created by Menlove and Healey [28] for the conterminous United States based on data from the USFS-FIA program. Given the consistent performance and proven success in AGB research, RF, Extreme Gradient Boosting (XGBoost), and ExtraTrees regressor were selected as the candidate ML models in our study.

Here, we investigate how temporally, and spatially discrete CFI data and aerial LIDAR can be combined with annual GSE data to develop ML models to predict forest AGB across seven states (MA, CT, VT, RI, NH, ME, NY) in northeastern United States. We hypothesize that Google Satellite Embeddings will improve the predictive accuracy and variance explained of machine learning-based aboveground biomass models and serve as a reliable complementary data source in areas with limited LiDAR coverage. To test this hypothesis, we aim to compare the performance of RF, XGBoost, and ExtraTrees models for AGB estimation through the integration of multi-year LiDAR data, AlphaEarth GSE, and NEFIN-CFI data across the northeastern United States. Our three-fold objectives are to; 1) develop a unifying framework that integrates spatially heterogeneous and unevenly distributed NEFIN-CFI data with multi-year LiDAR and GSE datasets for AGB estimation, 2). evaluate the contribution of LiDAR and GSE predictors in explaining AGB variability of NE-USA forests, and 3) compare the performance accuracy of RF, XGBoost, and ExtraTree models in AGB estimation across NE-USA forests. Our findings offer a new pathway for regional AGB modelling by combining aerial LiDAR, GSE and NEFIN-CFI data. We aim to address the gap in producing annual AGB maps at medium to fine spatial resolution by integrating GSE within this framework. Also, we expect to highlight the value of data-driven vegetation management and to provide an efficient AGB related information system to achieve regional carbon goals. Our findings will guide the selection of optimal RS data combinations, particularly in regions where airborne LiDAR is unavailable or less reliable. Finally, our models and findings will serve as a valuable and reproducible decision-making tool for landowners, forest managers, and policymakers who are engaged in regional forestry, vegetation, and carbon management decision making process.

## Method

### *Study Area*

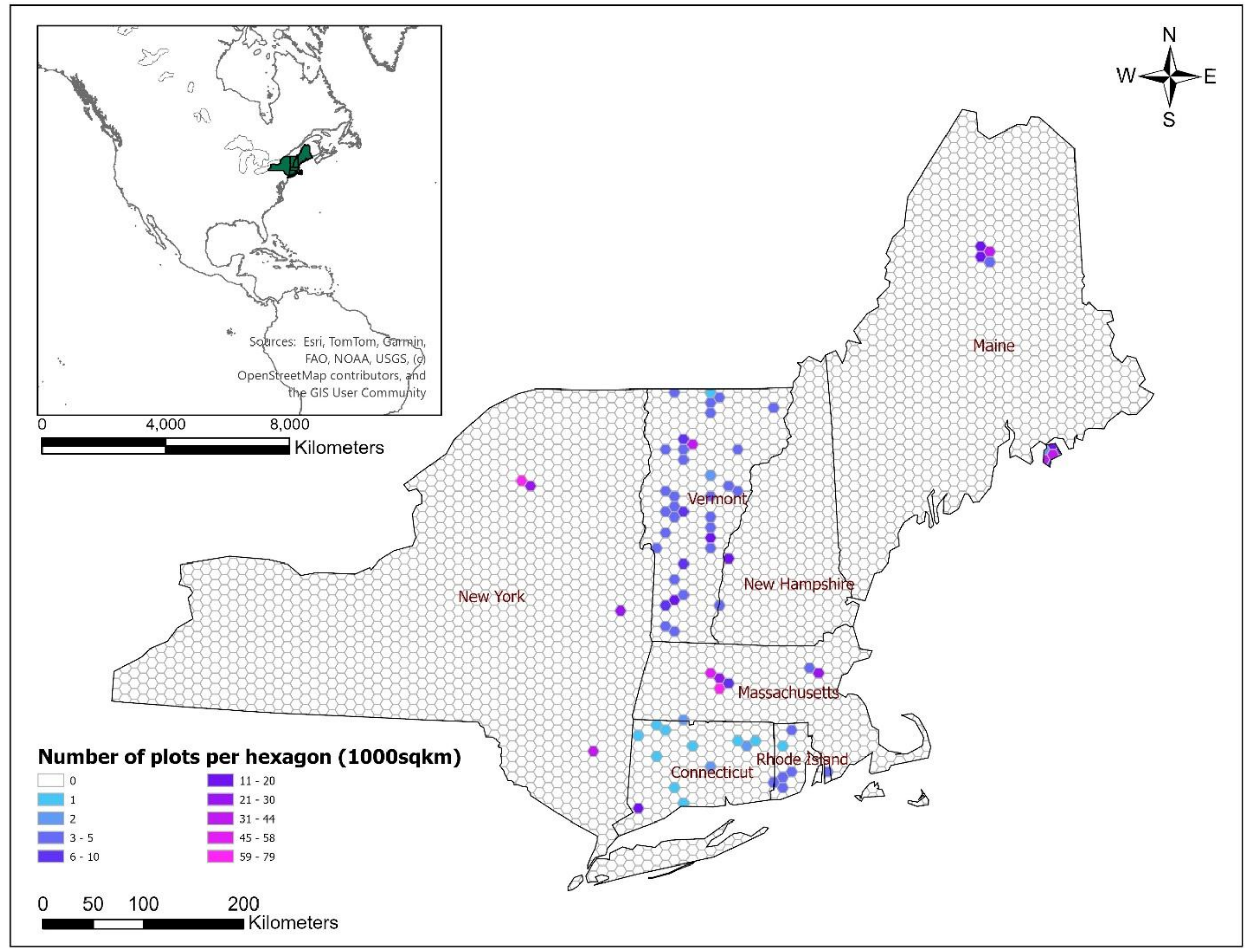

*Figure 1. Geographic distribution of continuous forest inventory plots used in our study across the Northeastern United States. Plot density has been summarized using 1000 km² hexagonal sampling units, with darker shades indicating higher plot densities.*

The study area (Figure 1) covers seven states of NEUSA, including Connecticut (CT), Massachusetts (MA), Maine (ME), New Hampshire (NH), New York (NY), Rhode Island (RI), and Vermont (VT). The study area extends approximately 40.5° - 47.5° N latitude and 80.0° - 66.5° W longitude covering more than 280 000 km$^2$ of the landscape [29]. The region generally has humid conditions with annual precipitation ranging from 914 mm to 1524 mm. Mean annual temperature ranges from 2.2°C to 13.9 °C [30]. The area is mostly covered by temperate forests, with evergreen and deciduous trees occupying more than 50% of the land [31]. The carbon sequestration rate in forests of the study area averages about 149 gC $m^{-2}$ $year^{-1}$ [32]. The area consisted of more than 114 tree species. The most abundant species by basal area are *Acer rubrum*, *Acer saccharum*, and *Pinus strobus* [33].

### *Northeastern Forest Inventory Network Data*

For several decades, the CFI programs have provided long-term records of forest conditions across the Northeastern United States. The NEFIN program aggregates and standardizes the CFI data from ten monitoring programs that are scattered across seven states of the northeastern region to one consolidated online tool. The database includes more than 5000 plot data reported since 1960 (Table 1) [10, 33]. We included only CFI programs that shared unfuzzed plot locations, which are presented in Table 1.

*Table 1. Continuous Forest Inventory (CFI) programs selected. Inventories containing fuzzed or intentionally displaced plot coordinates were excluded from our analysis.*

| Program | State | Duration | Status | Plot Shape | Plot size |
|---|---|---|---|---|---|
| Cadwell Mt Toby (CAD) | MA | 1983-2014 | Unfuzzed | Square | 14.14m |
| Northeast Temperate Network (NETN) | NEUSA | 2006-2024 | Unfuzzed | Square | 20.00m |
| Baxter State Park CFI | ME | 1999-2025 | Unfuzzed | Circular | 16.06m |
| State University of New York Forest Properties Continuous Forest Inventory (SUNY) | NY | 1970-2017 | Unfuzzed | Circular | 16.05m |
| Maine Ecological Reserves Program | ME | 2002-2024 | Unfuzzed | Circular | 16.06m |
| FEMC Forest Health Monitoring (FHM) | NH | 2021-2024 | Fuzzed | Circular | 07.32m |
| | VT | 1992-2021 | Unfuzzed | Circular | 07.32m |
| | ME | 2021-2024 | Fuzzed | Circular | 16.06m |
| | RI | 2021-2024 | UnFuzzed | Circular | 07.32m |
| | CT | 2021-2024 | UnFuzzed | Circular | 16.06m |
| | MA | 2019-2024 | Fuzzed | Circular | 16.06m |
| | NY | 2022-2024 | Fuzzed | Circular | 16.06m |

Most of the inventory programs within NEFIN use fixed radius plots with independent, clustered or nested plot designs. In contrast, NETN, CAD and SUNY datasets use square-shaped plots with fixed measurements [33]. In our study, we avoided using variable radius plots and plots with fuzzed plot locations [5, 11].

### *Response Variable Calculations*

#### *Aboveground Biomass Estimation*

We used updated allometric equations introduced by [34]. Chojnacky et al. [34] introduced 35 generalized equations using allometric scaling theory (Equation 1) compared to previous 10 equations introduced by Jenkins et al [9]. Plot-level AGB was calculated by adding individual tree estimates from Equation 1, for all trees with DBH ≥ 12.7 cm within each plot [5,35].

$$AGB = exp\ (\beta_0 + \beta_1\ ln\ (DBH)) \quad (1)$$

Where:

*AGB = aboveground biomass (in kg per individual tree)*

*DBH = diameter at breast height (cm)*

*$\beta_0$ and $\beta_1$ = species group specific coefficients*

*exp = exponential function*

*ln = natural logarithm*

#### *Annual Aboveground Biomass Adjustment*

The annual AGB adjustment (ΔAGB) was calculated by comparing plot level AGB between consecutive inventory measurements (Equation 2) of a given plot.

$$\Delta AGB = \frac{AGB_{t_2} - AGB_{t_1}}{t_2 - t_1} \quad (2)$$

Where, $AGB_{t_1}$ and $AGB_{t_2}$ represent the total plot-level AGB (kg) at the earlier and the most recent inventory year, respectively. The difference of time between measurements was denoted by $t_2$-$t_1$. This calculation yielded the mean annual rate of AGB accumulation (kg $yr^{-1}$). When the inventory measurements were collected within ±3 years of the LiDAR collection date, the ΔAGB × ($t_2$-$t_1$) was applied to adjust the plot-level AGB to correspond with AGB of the LiDAR acquisition year [11,36]. This step will ensure temporal consistency between field data and RS data in our modeling approach. Due to variation in plot area, AGB was standardized to $Mgha^{-1}$ by dividing plot biomass by plot area (ha) and converted from $kgha^{-1}$ to $Mgha^{-1}$ [11, 10]. These AGB values were served as response variables for the candidate ML models.

We followed several additional criteria proposed Lamahewage et al. [5] and Johnson et al. [11] to filter outlier plots before analysis. We excluded i). plots with AGB ≤0 $Mgha^{-1}$ with LiDAR mean height >1m, ii). plots that are not at least 90% covered by the corresponding LiDAR convex hull, iii). duplicate plots remeasured within ±3 years of the LiDAR acquisition, iv) fuzzed plot location and iv). plots outside of the NOAA C-CAP 1m and NLCD 30m (for VT) landcover masks [37,38].

***Remote Sensing Predictors***

*LiDAR Variables*

LiDAR data has been a key source of information on forest vertical structure in recent large-scale forest biometrics studies [39]. Our study aggregated and utilized more than 12 different LiDAR missions for the entire study region. We used most recent LiDAR missions for 10 years of available data (2011-2025) from USGS-3DEP LiDAR Explorer [10,40]. Both quality level 1 and 2 LiDAR data were acquired to retain the maximum coverage of the study area. LiDAR pulse densities ranged from 1-20 points/$m^2$ (Appendix 1). We used height normalized LiDAR data for modeling at 1m spatial resolution and were later resampled to 10-30 m grid cells as necessary [10]. Finally, a total of 29 independent LiDAR metrices were derived from LiDAR returns (Appendix 2) using ArcGIS Pro (version 3.5.4), and Python (version 3.11.9) "*PDAL, laspy* (version 2.7.0), and *rasterio*" packages [41]. We used the per-annum net growth adjustment to bring inventory year AGB equivalent to the AGB of LiDAR data collection year.

*Alpha Earth Google Satellite Embedding Dataset*

The AlphaEarth Foundations Satellite Embedding dataset is produced by Google and Google DeepMind. GSE dataset provides 64 analysis-ready features at 10 m spatial resolution [16]. We accessed the annual GSE collection in Google Earth Engine (ee.ImageCollection ("GOOGLE/SATELLITE_EMBEDDING/V1/ANNUAL")) and extracted values for each CFI plot across the available time series from 2017 to 2025. We computed summary statistics (maximum, mean, minimum, and standard deviation) for each of the 64 embedding features at the plot level. This resulted in 256 predictor variables (Appendix 2) per selected CFI location per year since 2017. We assumed that the 2017 embeddings adequately represent conditions during 2015-2017 to maximize the number of CFI plots retained for comparison with LiDAR-based models. This assumption is supported by the fact that the model is trained on multi-year observations [16].

***Principal Component Analysis***

First, we conducted a Pearson correlation analysis to visualize and to understand the correlation among the RS variables. Principal component analysis (PCA) was applied to the 256 GSE predictors prior to model training to reduce dimensionality and minimize redundancy of the feature space [42]. We standardized GSE predictors prior to PCA to ensure comparability among variables during dimensionality reduction via *"StandardScaler"* library of Python (v 3.11.9) [62]. PCA was subsequently implemented using the *"PCA"* function from the *"scikit-learn"* package of the Python (v 3.11.9). The number of retained components was determined based on a cumulative explained variance threshold of 95% [42]. The retained PCs were combined with the LiDAR and spatial variables to generate the final predictor dataset to be used subsequent ML model (RF, ExtraTrees, and XGBoost regression) training and testing.

***Spatial Autocorrelation Analysis***

We conducted a spatial autocorrelation analysis to assess underlying geographic structure within biomass modelling. We computed Moran's I index on ML model residuals as the first measure of spatial autocorrelation analysis [43]. Moran's I values were complemented with statistical significance assessed through permutation-based simulations under the null hypothesis ($p$ value $\leq 0.05$) of spatial randomness. Next, spatial covariates were established using spatial proxies such as plot center coordinates (latitude and longitude). Additional polynomial spatial features including squared latitude, squared longitude, and the latitude longitude interaction were derived [44, 45, 46]. The spatial derivatives that we used in our AGB models can be found in Appendix 2. These spatial derivatives were appended to the original predictor set or followed by PCA prior to model fitting [47]. Finally, Moran's I statistics were recalculated for the spatially corrected model residuals to assess whether incorporation of geographic covariates reduced residual spatial dependency across the study region [47, 43].

***Feature Importance Analysis***

We used permutation-based feature importance analysis to evaluate the contribution of predictor variables for each candidate ML model before and after the tuning process. We used 30 repeated permutations for each predictor, and 10-fold cross-validation (CV) in importance estimates [5, 15, 48]. The variables were then ranked according to their mean permutation importance values [19]. We only reported the top 20 ranked feature importance values for Scenario I features for comparison purposes.

***Machine Learning Model Training and Validation***

We categorized explanatory variables into three groups based on RS data source: i) G1: aerial LiDAR-only, ii). G2: GSE-only, and iii). G3: aerial LiDAR and GSE [5, 15, 48]. We trained and tested a sequence of ML models to compare the use of LiDAR and GSE variables under two different scenarios (Table 2). The scenario I models were trained with 589 growth adjusted plot locations for all 3 variable groups, where both LiDAR and GSE data are available. The scenario II was designed to evaluate the full potential of GSE dataset in addressing spatiotemporal discrepancy in inventory and LiDAR data collections. By only using GSE data and AGB growth adjustment, we expanded the sample size to 6,802 observations without masking the data out due to limited aerial LiDAR availability. This approach enabled a detailed comparison of GSE and LiDAR data without exponentially reducing the sample size due to mismatches in data collection years. Each model was fitted to a randomly selected 80% of the dataset leaving the remaining 20% for independent testing, with 10-fold CV applied for validation [11].

*Table 2. Plot selection criteria used under two scenarios to evaluate the effects of LiDAR availability and annual GSE observations on sample size and model performance.*

| **Scenario** | **Plot Inclusion Criteria** | **No. of plots** |
|---|---|---|
| Scenario I: Compare GSE and LIDAR | Plots were retained only if they overlapped the GSE observation period (2017-2025) and were inventoried within ±3 years of LiDAR acquisition. | 589 |
| Scenario II: maximum utility of GSE data | Plots were retained only if they were inventoried within ±3 years of the corresponding GSE data collection period (continuous 2017-2025). | 6801 |

*Random Forest (RF)*

The RF is one of the most used tree-based algorithms in forest AGB research. We used the Python version 3.11.9 environment, the *"scikit-learn"* library to implement the RF algorithm. We utilized both bootstrap aggregation (bagging) and random subset selection of variables at each split to reduce overfitting [10]. Model tuning, cross validation, and variable importance were conducted using *"GridSearchCV"* and *"KFold"* and built-in *"feature_importances_"* attributes respectively. We performed hyperparameter tuning via *GridSearchCV* (Python v 3.11.9) with a 10-fold shuffled CV with a fixed random seed of 42 [49]. The tested hyperparameter space is included in Table 3.

*Table 3. Hyperparameter ranges evaluated during grid-search optimization of the candidate machine learning algorithms used for aboveground biomass prediction.*

| Model | Hyperparameter | Search Space |
|---|---|---|
| Random Forest (RF) [5] | Number of trees (n_estimators) | 300, 500 |
| | Maximum tree depth (max_depth) | 10, 20, None |
| | Minimum samples required for node split (min_samples_split) | 2, 5 |
| | Minimum samples required at leaf node (min_samples_leaf) | 1, 2, 4 |
| | Number/proportion of variables considered at each split (max_features) | sqrt, 0.5, 1.0 |
| Extremely Randomized Trees (ExtraTrees) [51] | Number of trees (n_estimators) | 300, 500 |
| | Maximum tree depth (max_depth) | 10, 20, None |
| | Minimum samples required for node split (min_samples_split) | 2, 5 |
| | Minimum samples required at leaf node (min_samples_leaf) | 1, 2, 4 |
| | Number/proportion of variables considered at each split (max_features) | sqrt, 0.5, 1.0 |
| Extreme Gradient Boost (XGBoost) Tian et al. 2022 | Number of boosting trees (n_estimators) | 300, 500 |
| | Maximum tree depth (max_depth) | 3, 5, 7 |
| | Learning rate (learning_rate) | 0.01, 0.05, 0.1 |
| | Row subsampling ratio (subsample) | 0.7, 0.9 |
| | Column subsampling ratio (colsample_bytree) | 0.7, 0.9 |
| | L2 regularization term (reg_lambda) | 1, 5, 10 |

*Extreme Gradient Boosting (XGBoost)*

The XGBoost is another tree-based ML algorithm that often performs well with complex datasets. The algorithm builds trees sequentially, with each new tree fitted to reduce the residual errors from the previous trees [50]. We used the *XGBRegressor* function from the "*xgboost*" python (3.11.9) library. The models were configured with a squared error regression objective and a fixed random seed of 42 [50]. Hyperparameter optimization was conducted as mentioned in Table 3 using *"GridSearchCV*".

*Extremely Randomized Trees (ExtraTrees)*

The ExtraTrees algorithm constructs highly randomized decision trees [51]. Here, split thresholds are selected randomly, reducing model variance and helping to limit overfitting in small datasets. We implemented the ExtraTrees algorithm using *"ExtraTreesRegressor"* from the *"scikit-learn"* ensemble module of python version 3.11.9. The ExtraTrees algorithm introduces additional randomization in the selection of splits to reduce variance and improve generalization. We performed the hyperparameter tuning using *"GridSearchCV"* with 10-fold CV as for previous candidate models with random seed of 42 [25, 51]. The tested hyperparameter space is included the Table 3.

***Model Performance Evaluation***

We compared the model performance against 20% of hold-out testing data for each ML model. The accuracy metrics that we used are presented in Equations 3-7: i). root mean square error (RMSE), ii). Mean absolute error (MAE), iii) coefficient of determination ($R^2$), v). Bias and vi) relative bias (RBias). Where $n$ is the number of plots, $\hat{y}_i$ is the predicted AGB, $y_i$ is the Chojnacky et al. [8] estimated AGB of the corresponding location, and $\bar{y}$ is the mean field AGB estimation of all the plots included in the testing dataset [5, 10, 11].

$$\text{RMSE}=\sqrt{\frac{1}{n}\sum_{i=1}^{n}(y_i-\hat{y}_i)^2} \tag{3}$$

$$\text{MAE}=\frac{1}{n}\sum_{i=1}^{n}(|y_i-\hat{y}_i|) \tag{4}$$

$$R^2=1-\frac{\sum_{i=1}^{n}(y_i-\hat{y}_i)^2}{\sum_{i=1}^{n}(y_i-\bar{y}_i)^2} \tag{5}$$

$$\text{Bias}=\frac{1}{n}\sum_{i=1}^{n}(\hat{y}_i-y_i) \tag{6}$$

$$\text{Rbias}=\left(\frac{\text{Bias}}{\bar{y}}\right)100\% \tag{7}$$

***Monte- Carlo Simulation for Algorithm Stability and Variable Sensitivity Propagation***

We conducted Monte Carlo model stability and permutation-based variable group sensitivity analyses to understand the ML model stability and predictor group dependency among the ML models. In both analyses, the Python packages "*NumPy*" and "*Pandas*" were employed for permutation procedures and data management operations [52]. We compared six candidate ML models developed from the combined LiDAR and GSE variable group (G3), before and after hyperparameter tuning. We simulated 100 iterations using varying random seeds and distributions of performance metrics ($R^2$, RMSE, and MAE) across iterations for Monte Carlo model stability assessment. Next, predictor variables were grouped into seven categories representing all possible variable combinations of LiDAR, GSE-PCs selected with 95% variance (GSE_PCA95), and spatial predictors. Sensitivity was quantified as reductions in predictive performance ($\Delta R^2$) and increases in prediction error ($\Delta$RMSE) relative to baseline model performance during 100 simulations [53].

## Results

### *Feature Correlation*

Correlation analysis revealed weak to moderate relationships between LiDAR variables and GSE features, with most correlation coefficients falling below ±0.65. Upper height percentiles, rugosity, and canopy standard deviation exhibited the strongest coefficients with some embedding features (i.e. A28_mean and A36_mean) ranging from ±0.24 to ± 0.65. However, LiDAR density metrics and height bins generally showed weak correlations with embedding derivatives with most values remaining ≤ 0.10. Strong multicollinearity was observed among some LiDAR variables themselves, especially among upper canopy percentiles and roughness values. The detailed correlation heatmap is provided in Appendix 3.

### *PCA for embedding variables*

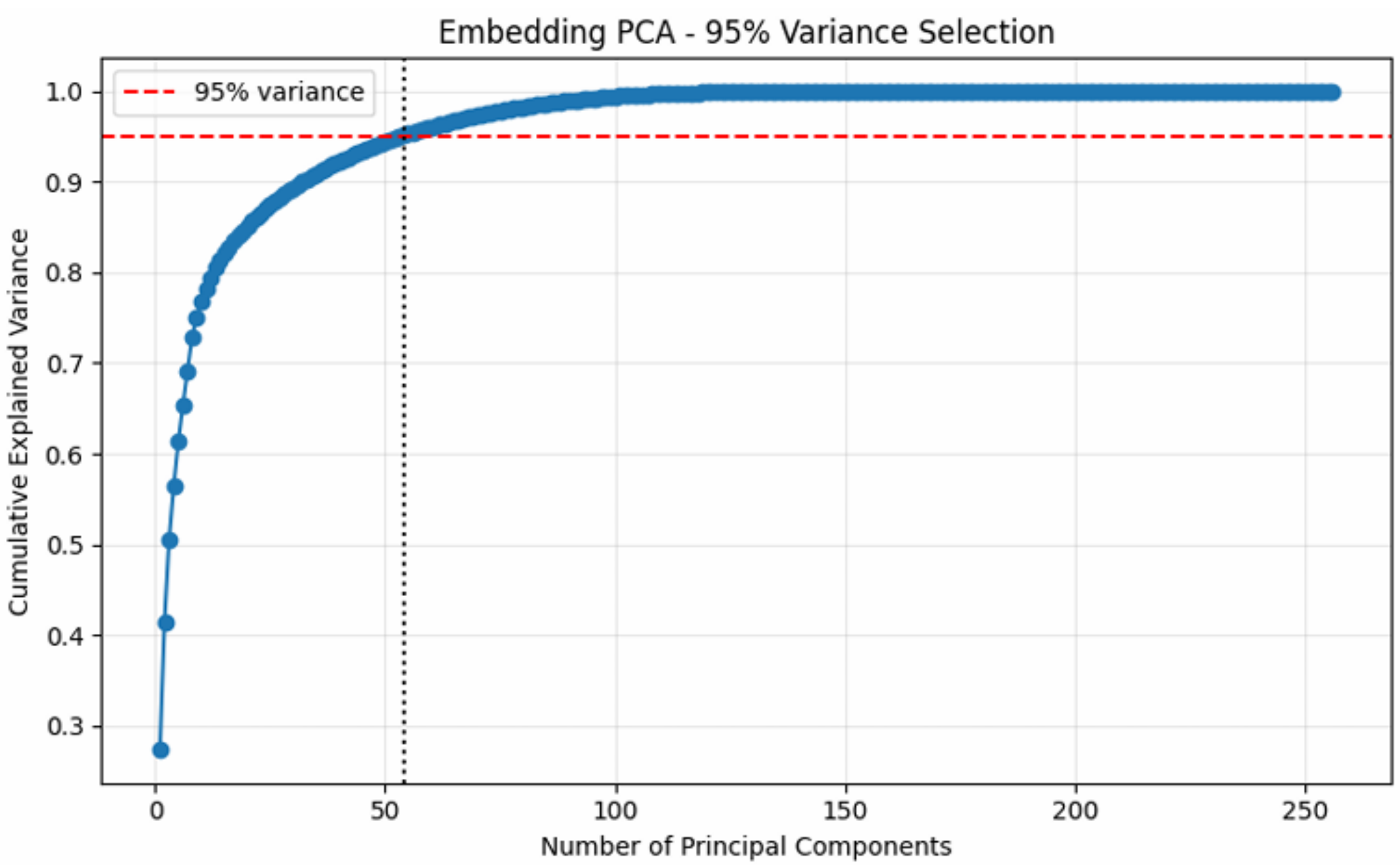

*Figure 2. Cumulative variance explained (Y axis) by principal components (X axis) derived from Google Satellite Embeddings. Fifty-three principal components were required to explain 95% of the total variance and the red dashed line denotes the 95% variance threshold.*

We conducted the principal component analysis on the 256 AlphaEarth GSE predictors (Appendix 2). PCA substantially reduced the dimensionality of the embedding space while retaining most of the variance structure present within the original dataset. The cumulative explained variance curve (Figure 2) demonstrated a rapid increase in explained variance within the first 9 PCs. The first 32 PCs explained 90% of the variance, while 54 and 92 PCs captured 95% and 99% of the variance, respectively. The reduction from 256 original predictors to only 54 PCs with preserving 95% of variance indicates that the AlphaEarth GSE were highly structured rather than independent.

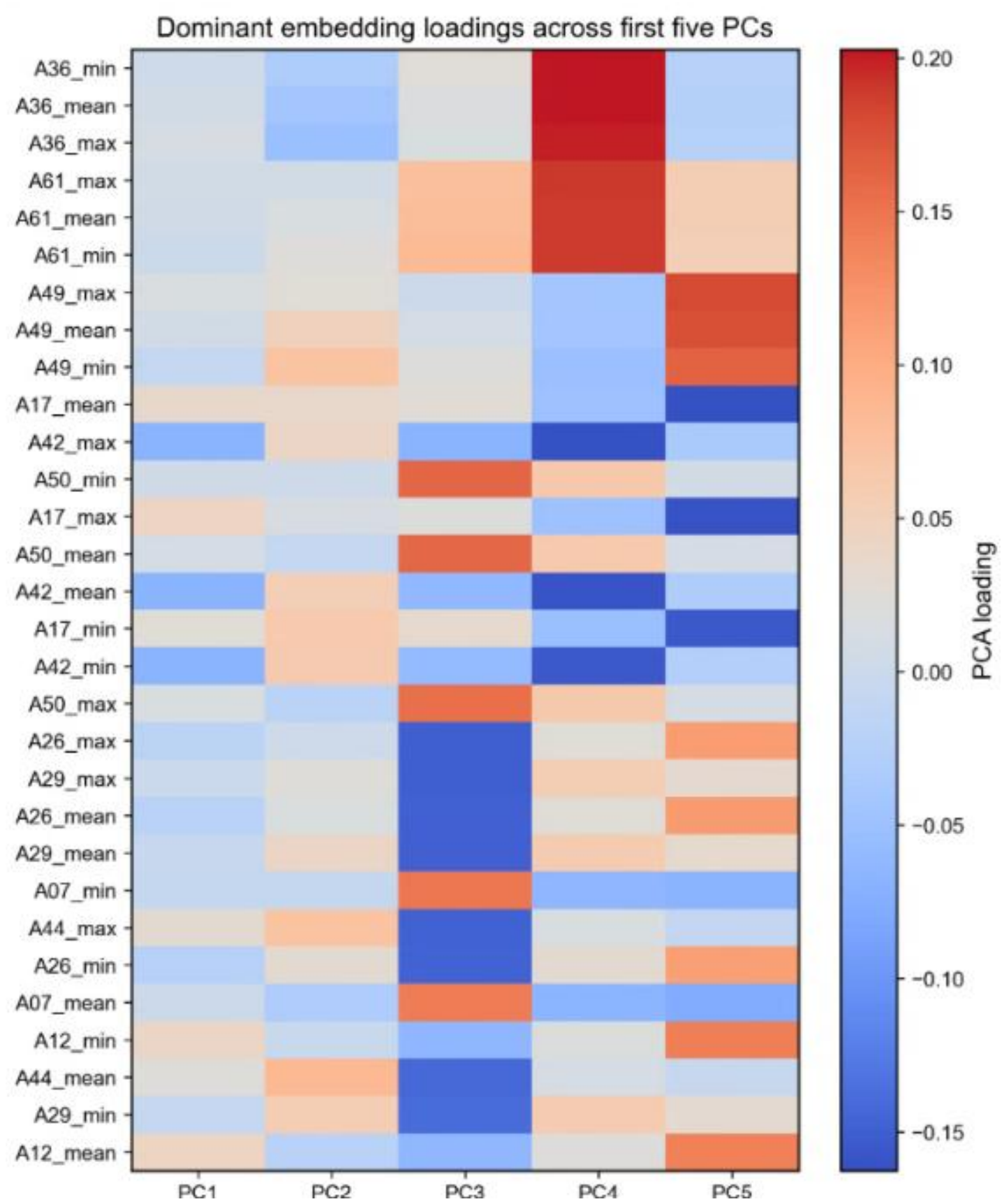

*Figure 3. Principle component analysis (PCA) loading heatmap. This figure illustrates the contribution of Google Satellite Embedding features (Y axis) to the first five principal components (X axis). Red colors indicate positive loadings, whereas blue colors indicate negative loadings. The magnitude of the loading reflects the relative contribution of each embedding feature to the corresponding principal component.*

In PCA loading heatmap (Figure 3), positive loading values indicate that GSE variable and PC scores have a positive relationship, whereas negative loading values decrease the principal component score. GSE features A36, A61, A49, A17, A42, A50, A26, and A29 displayed comparatively stronger contributions to the first five PCs. Positive and negative loadings within the same principal component possibly indicate that the PCA separated contrasting vegetation and canopy characteristics. For example, PC4 showed strong positive loadings associated with A36 and A61 features, while A42 exhibited strong negative loadings. However, due to the black box nature of embedding features and PC layers we are unable to identify exactly which ecological pattern that caused the pattern.

***Treating Spatial Autocorrelation***

Spatial autocorrelation analyses significantly improved the spatial independence of model residuals (Table 4). Prior to spatial correction, the AGB models exclusively used LiDAR data (G1) exhibited the highest residual clustering, with Moran's I values ranging from 0.0930 to 0.2471 (p value<0.05). This indicated that prediction errors were geographically structured when spatial predictors were excluded.

*Table 4: Moran's I statistics and associated p-values calculated from model residuals before and after spatial correction. Moran's I values quantify the degree of residual spatial autocorrelation, while p-values indicate the statistical significance of the observed spatial pattern.*

| **RS variable Group** | **Model** | **Moran's I (Before)** | **p-value** | **Moran's I (After)** | **p-value** |
|---|---|---|---|---|---|
| G1: LiDAR-only | RF | 0.2305 | 0.001 | 0.0698 | 0.033 |
| | ExtraTrees | 0.2471 | 0.001 | 0.0483 | 0.088 |
| | XGBoost | 0.0930 | 0.009 | 0.0464 | 0.090 |
| G2: GSE-only | RF | 0.1246 | 0.004 | 0.0827 | 0.023 |
| | ExtraTrees | 0.0385 | 0.137 | 0.0714 | 0.350 |
| | XGBoost | 0.0686 | 0.051 | 0.0251 | 0.202 |
| G3: LiDAR and GSE | RF | 0.0809 | 0.034 | 0.0352 | 0.132 |
| | ExtraTrees | 0.0351 | 0.141 | 0.0344 | 0.138 |
| | XGBoost | 0.0342 | 0.147 | -0.0095 | 0.482 |

Residual spatial autocorrelation varied substantially among predictor groups and declined consistently following spatial correction (Table 2). Incorporation of spatial variables markedly reduced autocorrelation across all LiDAR models, decreasing Moran's I values by approximately 50-80%, with residual clustering becoming statistically nonsignificant for the ExtraTrees and XGBoost models. The GSE-only models (G2) showed comparatively lower initial spatial autocorrelation, suggesting that the embedding variables inherently captured portions of the regional spatial structure. Following spatial correction, Moran's I values remained relatively low (0.0251-0.0827), with XGBoost exhibiting the lowest Moran's I value of 0.0251 ($p > 0.05$). Overall, the AGB models produced using both LiDAR and GSE (G3) variables showed the weakest residual spatial autocorrelation. Moran's I values remained below 0.0809 in all combined models and approached zero in the XGBoost model after incorporating spatial predictors ($I = -0.0095$, $p > 0.05$). This suggests that combining LiDAR and GSE features successfully accounted for most of the spatial patterns present in the biomass data.

***Feature Importance***

Variable importance patterns varied substantially between the non-spatial and spatially corrected candidate models across all predictor groups. Canopy structural variables, such as roughness, canopy relief ratio, upper height percentiles, and height bin metrics consistently ranked among the most influential predictors in the non-spatial LiDAR-only models (G1). As expected, explaining vertical canopy complexity and structural heterogeneity were the primary controls on AGB estimation. However, following spatial correction, geographic variables including latitude derivatives, distance, and trigonometric spatial transformations became dominant across all algorithms, particularly in the RF and XGBoost models (Appendix 4.a).

The GSE-only models (G2) exhibited noticeably different variable importance patterns. A relatively small subset of GSE derivatives dominated the non-spatial models (i.e. A19_max, A37_mean, A29_max). These variables likely represent underlying ecological gradients embedded within the AlphaEarth GSE feature space. After PCA dimensionality reduction and spatial correction, PCs such as PC23, PC28, and PC37 consistently identified as dominant predictors. This indicates that PCA components were already capturing large portions of the ecological and spatial variability embedded in the AlphaEarth GSE features (Appendix 4.b).

The ML models (G3) integrated both LiDAR and GSE derived metrics resulted in more balanced predictor importance distributions. For non-spatial ML models GSE metrics such as A37_mean and A50_min consistently appeared in each candidate model. In contrast, roughness, bin_20_25, crr and p95 appeared in at least two models and ranked at the highest important variables. The RF and the XGBoost models relied more heavily on GSE variables showing 1:3 ratio with LiDAR metrics, whereas ExtraTrees showed a more balanced LiDAR:GSE contribution with 2:3 ratio. The tuned spatial models increasingly prioritized spatial predictors alongside selected PCs and LiDAR variables. Especially, spatial predictors overwhelmingly dominated model importance in GSE-only (G3) XGBoost model. These results demonstrate that spatial correction re-adjusted the variable importance ranking from purely structural canopy metrics toward broader geographic variables. This substantially improved the spatial robustness and generalization of biomass predictions. Which highlights the requirement of well-distributed, ecological and geographically representative sampling designs for large area forest AGB modeling (Appendix 4.c).

### *Machine Learning Model Performance*

*Scenario I Results*

We observed significant differences in predictive performance among the three predictor groups and scenario I candidate ML models. Table 5 presents the variation in predictive performance among ML algorithms under different treatment strategies. Overall, model performance in relation to testing RMSE, MAE, Bias, Rbias, and $R^2$ values improved considerably following spatial correction and hyperparameter tuning.

*Table 5. Predictive performance of candidate machine learning models three predictor variable groups (LiDAR-only, GSE-only, and combined LiDAR and GSE).*

| RS Group | Model | No. of Variables | Model Name | Model Treatment | $R^2$ | RMSE/ $Mgha^{-1}$ | MAE/ $Mgha^{-1}$ | Bias/ $Mgha^{-1}$ | RBias |
|---|---|---|---|---|---|---|---|---|---|
| G1: LiDAR only | RF | 29 | RF1_D | Default Settings | 0.35 | 705.98 | 471.61 | 134.96 | 20.72% |
| | ExtraTrees | 29 | ET1_D | | 0.49 | 626.50 | 424.19 | 132.75 | 20.38% |
| | XGBoost | 29 | XG1_D | | 0.23 | 768.70 | 487.86 | 96.89 | 14.87% |
| | RF | 45 | RF1_S | Default Settings with Spatial Correction | 0.76 | 433.56 | 250.92 | 82.64 | 12.69% |
| | ExtraTrees | 45 | ET1_S | | 0.77 | 421.86 | 246.20 | 61.97 | 9.51% |
| | XGBoost | 45 | XG1_S | | 0.77 | 425.02 | 242.17 | 74.54 | 11.44% |
| | RF | 45 | RF1_T | Tuned Settings with Spatial Correction | 0.79 | 401.59 | 225.40 | 61.39 | 9.42% |
| | ExtraTrees | 45 | ET1_T | | 0.78 | 413.01 | 230.06 | 58.37 | 8.96% |
| | XGBoost | 45 | XG1_T | | 0.78 | 409.10 | 242.27 | 66.99 | 10.28% |
| G2: GSE only | RF | 256 | RF2_D | Default Settings | 0.54 | 602.90 | 396.22 | 133.02 | 19.84% |
| | ExtraTrees | 256 | ET2_D | | 0.62 | 549.86 | 355.73 | 125.16 | 18.66% |
| | XGBoost | 256 | XG2_D | | 0.53 | 613.27 | 382.16 | 100.06 | 14.92% |
| | RF | 70 | RF2_S | Default Settings with PCA and Spatial Correction | 0.64 | 530.08 | 299.63 | 87.44 | 13.61% |
| | ExtraTrees | 70 | ET2_S | | 0.59 | 562.00 | 318.36 | 109.71 | 17.08% |
| | XGBoost | 70 | XG2_S | | 0.60 | 547.03 | 308.44 | 71.92 | 11.20% |
| | RF | 70 | RF2_T | Tuned Settings with PCA and Spatial Correction | 0.64 | 524.32 | 297.75 | 87.32 | 13.60% |
| | ExtraTrees | 70 | ET2_T | | 0.66 | 509.59 | 288.83 | 77.06 | 12.00% |
| | XGBoost | 70 | XG2_T | | 0.67 | 503.02 | 292.13 | 73.07 | 11.37% |
| G3: LiDAR and GSE | RF | 285 | RF3_D | Default Settings | 0.54 | 595.42 | 383.23 | 145.20 | 22.29% |
| | ExtraTrees | 285 | ET3_D | | 0.71 | 476.72 | 282.85 | 106.12 | 16.29% |
| | XGBoost | 285 | XG3_D | | 0.47 | 637.73 | 368.45 | 71.72 | 11.01% |
| | RF | 99 | RF3_S | Default Settings with LiDAR, PCA, and Spatial Correction | 0.76 | 428.69 | 245.14 | 91.12 | 13.99% |
| | ExtraTrees | 99 | ET3_S | | 0.78 | 413.32 | 234.44 | 73.44 | 11.26% |
| | XGBoost | 99 | XG3_S | | 0.63 | 532.80 | 291.29 | 144.71 | 22.21% |
| | RF | 99 | RF3_T | Tuned Settings with LiDAR, PCA, and Spatial Correction | 0.77 | 419.09 | 237.20 | 73.98 | 11.36% |
| | ExtraTrees | 99 | ET3_T | | 0.77 | 418.29 | 228.65 | 67.48 | 10.35% |
| | XGBoost | 99 | XG3_T | | 0.79 | 402.09 | 237.05 | 75.01 | 11.52% |

Among the LiDAR-only (G1) models, the default ExtraTrees model (ET1_D) produced the strongest predictive performance with an $R^2$ of 0.49 and an MAE of 424.19 $Mgha^{-1}$ outperforming both RF and XGBoost counterparts. Incorporating spatial variables evidently improved the accuracy of all three models. The spatially corrected ExtraTrees model (ET1_S) and XGBoost spatial (XG1_S) achieved the highest $R^2$ of 0.77. However, we observed the lowest MAE (421.86 $Mgha^{-1}$) and RBias of 9.51% for ET1_S suggesting better overall prediction stability and lower systematic error. Hyperparameter tuning marginally increased the $R^2$ values of the three ML models by less than 4% compared to spatial models. But the RF1_T model achieved the best $R^2$ (0.79) and lowest RMSE and MAE values showing more than 25% of reduction in Bias and Rbias values. Compared with ExtraTrees and XGBoost, the RF algorithm benefited the most from tuning while improving model generalization and reducing systematic over or under prediction.

Generally, GSE-only (G2) default models performed stronger than the LiDAR-only (G1) default models. The ExtraTrees algorithm showed the best testing accuracies with a $R^2$ of 0.62, MAE of 355.73 $Mgha^{-1}$ and RBias of 18.66%. However, GSE-only spatial models performed inferior to LiDAR-only spatial models (Table 5). Even, the $R^2$ value for ET2_S model was reduced by 4.84% and $R^2$ of XG2_S increased marginally from 0.53 to 0.60. Likely the reason for that could be, PCA prioritizes components explaining maximum variance in the predictor dataset rather than components most strongly related to biomass [54]. Despite fluctuations in $R^2$ values, accuracy metrics such as RMSE, MAE, Bias, and RBias improved by 10-30% across all spatial ML models in GSE-only (G2) category. These trends indicate that PCA and spatial treatment enhanced model stability by minimizing redundancy and multicollinearity among predictors [42]. The tuned XGBoost model (XG2_T) statistically outperformed all other GSE models, producing the highest $R^2$ (0.67), lowest MAE (292.13 $Mgha^{-1}$), and lowest RBias (11.37%) values. However, the tuned GSE-only models (G2) displayed lower predictive performance compared to the tuned LiDAR-only model (G1) equivalents. The best LiDAR-only model (RF1_T) achieved an $R^2$ of 0.79 and MAE of 225.40 $Mgha^{-1}$ whereas the best GSE model (XG2_T) only achieved an $R^2$ of 0.67 and MAE of 292.13 $Mgha^{-1}$. This signifies approximately a 15.2% lower explanatory power and a 29.6% increase in prediction error (Figure 4).

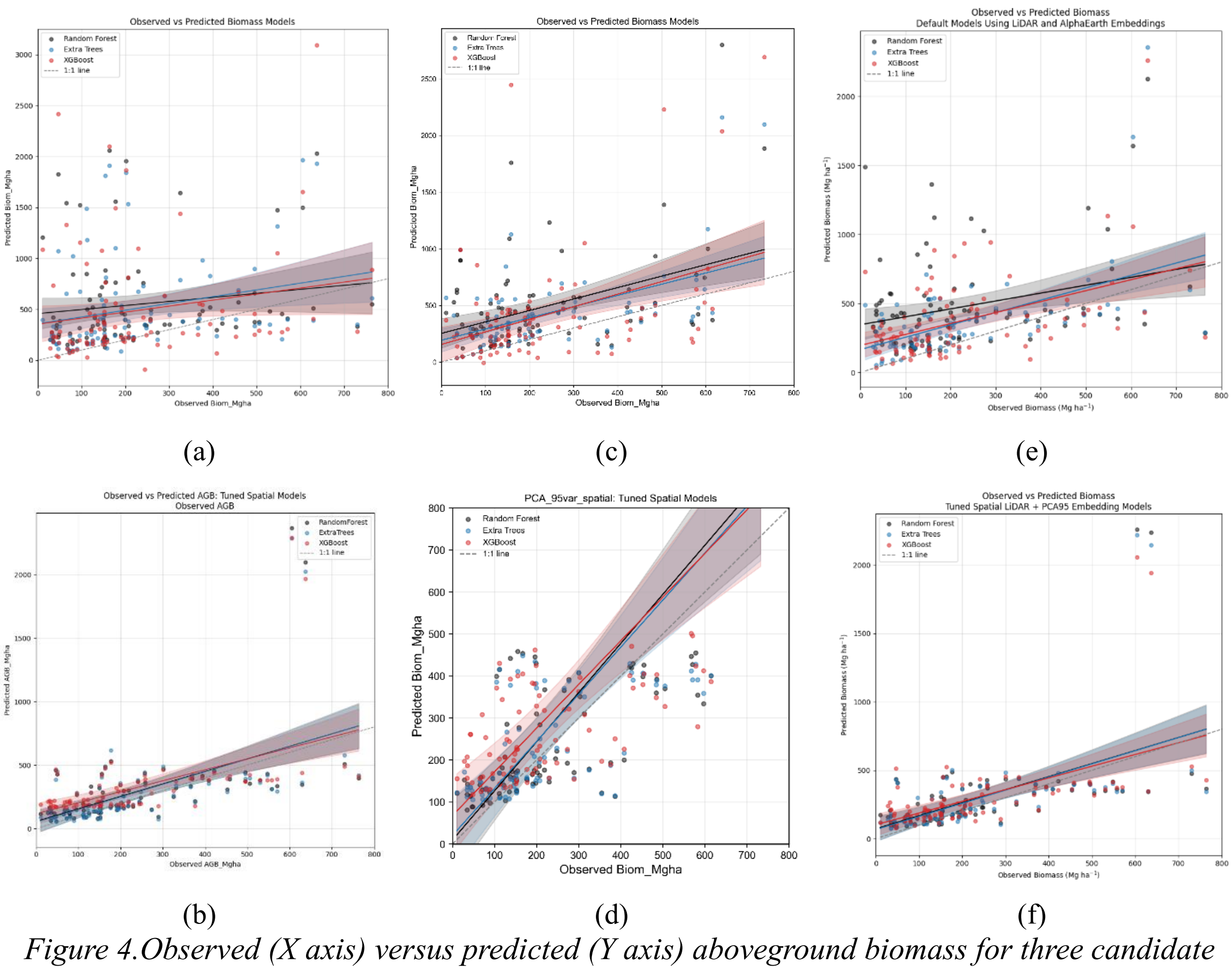

*Figure 4.Observed (X axis) versus predicted (Y axis) aboveground biomass for three candidate models under different predictor groups and hyperparameter tuning. Panels represent; LiDAR-only models (a). default, (b) tuned, GSE-only models (c) default, (d)tuned, and combined LiDAR-GSE models (e) default and (f)tuned. The dashed line indicates the 1:1 relationship, while shaded areas represent 95% confidence intervals around model predictions.*

The last group combined both LiDAR and GSE variables (G3) for model training and testing. The integration of both predictor types yielded the strongest overall model performance. Under default settings, ExtraTrees (ET3_D) substantially outperformed RF and XGBoost counterparts, achieving $R^2$ of 0.71 and MAE of 282.85 $Mgha^{-1}$. Among spatially corrected models, ET3_S achieved the best statistical performance with the highest $R^2$ of 0.78 and lowest MAE of 234.44 $Mgha^{-1}$. Hyperparameter tuning produced the best overall results of the study. Both RF3_T and XG3_T achieved $R^2$ of 0.79, while ET3_T maintained the lowest MAE values of 227.81 $Mgha^{-1}$, RBias of 10.72% and 0.78 of $R^2$. Statistically, the tuned XGBoost model (XG3_T) using combined LiDAR and GSE predictors represented the optimal modeling configuration across all experiments. Notably, for the group combined LiDAR and GSE variables, hyperparameter tuning resulted in only marginal improvements after the PCA and spatial correction was applied (Figure 5). The selected hyperparameter configurations are mentioned in Appendix 5. a. Overall, the combined LiDAR and GSE predictor framework consistently outperformed single-source predictor groups in terms of model performance and spatial dependency.

*Scenario II Results*

In Scenario II, annual GSE observations enabled us to expand the dataset from 589 to 6801 plots by propagating growth adjusted AGB estimates since 2017. This resulted in a nearly tenfold increase in sample size. The hyperparameter tuned XGBoost model with PCA-derived GSE features and spatial predictors achieved the highest predictive accuracy exhibiting an $R^2$ of 0.82 and an MAE of 245.23 Mg $ha^{-1}$ yielding the highest $R^2$ observed among the two scenarios. Detailed hyperparameter grids and model performance metrics are provided in Appendix 5.b.

***Monte Carlo Simulation for Algorithm Stability***

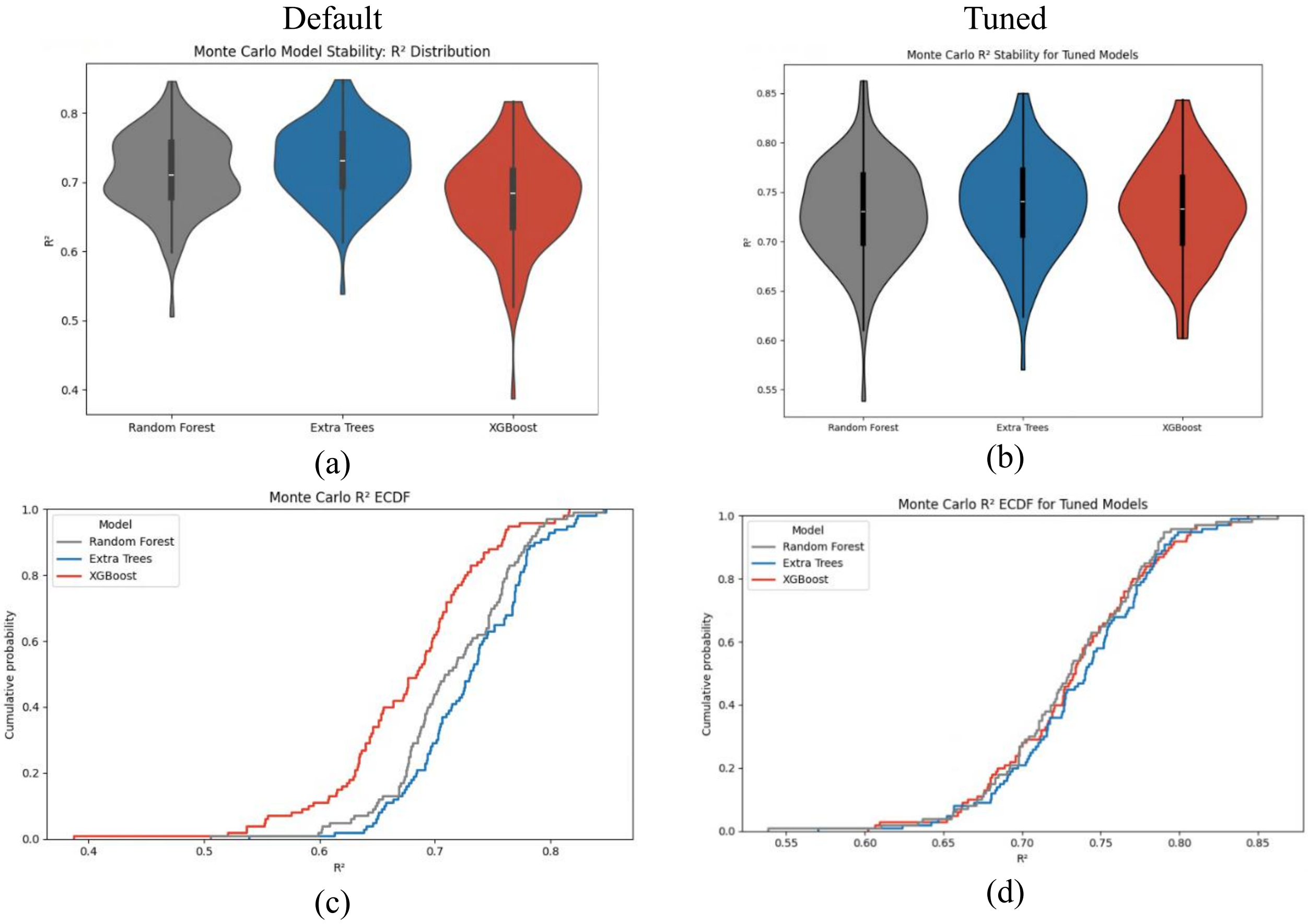

*Figure 5. Monte Carlo assessed model stability for machine learning models trained using combined LiDAR and GSE predictors. Violin plots summarize the distribution of $R^2$ values across 100 Monte Carlo iterations under default (a) and hyperparameter-tuned (b) settings. Corresponding empirical cumulative distribution functions (ECDFs) are shown for default (c) and tuned (d) models. Together, these plots illustrate the model performance under repeated random sampling.*

According to the Monte Carlo stability statistics and empirical cumulative distribution function (ECDF) plots, Hyperparameter optimization improved both predictive performance and algorithm stability of all candidate ML models. Among the default models, ExtraTrees exhibited the strongest predictive performance, with the highest mean Monte Carlo $R^2$ of 0.73. The XGBoost model demonstrated the largest variability across Monte Carlo iterations showing standard deviation of $R^2 = 0.07$. Following hyperparameter optimization, all three algorithms have improved predictive consistency showing reductions in the standard deviation of Monte

Carlo $R^2$ values ranging from 9.40% to 27.90%. However, the ExtraTrees model exhibited only a marginal increment of Monte Carlo $R^2$ from 0.73 to 0.74 after hyperparameter optimization. The narrower $R^2$ distributions and steeper ECDF curves observed after tuning indicated that hyperparameter optimization reduced performance variability, thereby improving general stability. According to Figure 5, the tuned XGBoost model demonstrated the greatest improvement with a narrower $R^2$ distribution and reduced lower-tail variability. Among the three candidate algorithms, ExtraTrees consistently demonstrated the most stable performance across Monte Carlo iterations.

### ***Permutation Based Variable Group Sensitivity***

The variable sensitivity analysis revealed that predictor dependence varied substantially among ML algorithms (Figure 6). GSE-derived PCA variables exhibited the least sensitivity with mean $\Delta R^2$ ranging from 0.001-0.014. In contrast, the strongest sensitivity responses were observed from combining LiDAR, GSE, and spatial variables. This combination generated the largest mean $\Delta R^2$ reductions across all models, reaching 0.23 for the RF model, 0.21 for ExtraTrees, and 0.26 for XGBoost. Similarly, LiDAR and spatial combinations remained highly influential with mean $\Delta R^2$ of 0.19-0.21. These trends explain that structural canopy metrics and geographic context collectively represent the primary drivers of AGB estimation performance. The mean $\Delta R^2$ values were increased to 0.05 for RF, 0.11 ExtraTrees, and 0.16 for XGBoost when GSE PCs were integrated with spatial information. Although GSE predictors did not fully replace LiDAR structural information under the present framework, integrating GSE with spatial predictors offers a promising alternative where LiDAR data are unavailable.

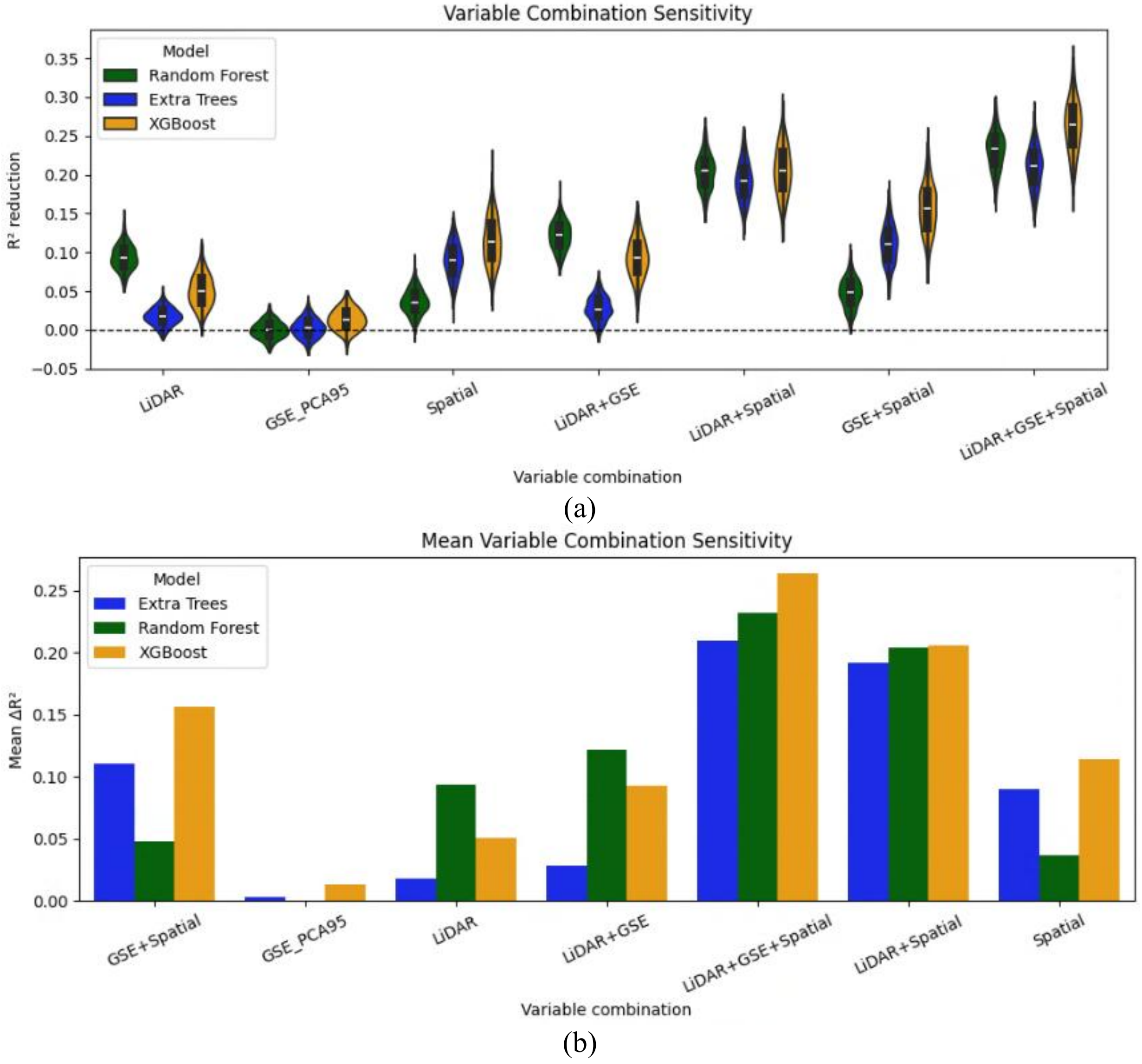

*Figure 6. Permutation-based variable group sensitivity analysis for machine learning models trained using different predictor combinations. (a) Violin plots showing the distribution of reductions in $R^2$ ($\Delta R^2$) different predictor group combinations across 100 iterations. (b) Mean $\Delta R^2$ values summarizing the relative contribution of each predictor group to model performance. Higher $\Delta R^2$ values indicate greater importance of the corresponding variable group for aboveground biomass prediction.*

**Discussion**

For decades, researchers sought to develop accurate AGB and forest carbon models across multiple spatial scales. In recent years, AGB estimation has increasingly shifted toward spatially explicit high resolution modeling approaches with advancements in AI and RS technologies. Most of the current AI based AGB research reached 0.20-0.65 of $R^2$ values with regional or national forest inventory data [5, 10, 22]. However, some researchers were able to achieve higher percentage of variance explained via integrating multimodal RS data, high accurate field data including destructive sampling. As instances, Xu et al. [14], Nandy et al. [15], and Sheridan et al. [7] obtained higher $R^2$ values respectively, 0.69, 0.87, and 0.83. In this study, we achieved $R^2$ of 0.67-0.79 using GSE features, LiDAR and ML algorithms indicating higher percentage variance explained for the region. We also observed that using the full GSE time series without restricting plots with availability of LiDAR coverage increased $R^2$ from 0.67 to 0.82 for northeastern U.S. forests. These results demonstrate that AGB model performance varied considerably among RS data source, sample size, algorithms, and spatial treatments.

The improved performance observed in Scenario II likely resulted from the substantial increase in sample size enabled by annual GSE observations. This increased sample size by incorporating plots previously excluded due to limited LiDAR coverage and improved representation of environmental variability across the region [54]. These results ($R^2$ =0.82) suggest that annual GSE observations can potentially overcome limitations associated with spatially and temporally discrete LiDAR coverage by substantially increasing the number of training samples available for AGB modeling.

The high RMSE and MAE values observed in our study are likely associated with the inclusion of extremely high AGB plots (>600 $Mgha^{-1}$), which were not statistical outliers and corresponded with extremely high LiDAR mean heights. Such high AGB conditions inherently increased model uncertainty due to under or over prediction at the extremes. Similar challenges have been reported in previous forest biomass studies where increased variance occurs at high biomass ranges [6, 55]. As seen in Figure 7, the residual distributions of the tuned models closely aligned with the 1:1 line where AGB <300 $Mhha^{-1}$ despite the relatively high MAE values observed due to the occurrence of extreme AGB conditions. These unlikely high AGB values could result from allometric uncertainty, extreme AGB conditions, catalyzed annual growth adjustments due to disturbance, silvicultural treatment, or potential geolocation mismatches between field inventory plots and remotely sensed data [56]. Also, spatially clustered and different small plot sizes could cause these extreme AGB values given the associated uncertainties in calculating AGB density [57]. According to Winsemius et al. [57] larger and spatially representative inventory plots generally reduce edge effects and improve alignment with RS observations in AGB estimation. Here, we used NEFIN data collected from ten different CFI programs with different plot sizes from 7 m-20 m (Table 1), which added another uncertainty to field AGB estimates. According to Vorster et al. [56] explained that allometric uncertainty contributes to 30-75% of total uncertainty based on the equation. We used updated generalized allometric equations for the North American tree species introduced by Chojnacky et al. [8]. This approach resulted in 35 generalized equations compared to the 10 equations that were derived by Jenkins et al. [9]. However, both approaches generally overestimate AGB by approximately 20% compared to FIA component ratio equations [8]. However, allometric model uncertainty was not evaluated in our study.

All LiDAR datasets used in our study varied in pulse densities from 1-20 points $m^{-2}$, representing both quality level 1 and 2 acquisitions. In addition, field AGB values were adjusted using annual growth corrections only including the plots measured within ±3 years from LiDAR acquisition year [57]. Also, plot exclusion criteria described in Lamahewage et al. [5] and Johnson et al. [11] were applied to remove obvious spatial outliers and to eliminate remeasured plots. These treatments reduced the sample size from 5000 to 589 given the compatibility with spatial and temporal distribution of LiDAR coverage. However, these CFI datasets were collected under different program objectives and sampling frameworks, often with unknown geolocation uncertainties that may exceed 30 m [33]. Some plot locations may have undergone through the silvicultural treatments or unreported disturbances further explaining extreme high and low AGB values. Potential solutions include increasing geographic representativeness and improving the geolocation accuracy of forest inventory networks in future studies.

One of our key objectives was to understand the utility of GSE dataset as a substitute for where LiDAR data is unavailable. Also, GSE is a global dataset with 10m resolution which is produced annually. This characteristics are highly valuable towards generating the annual AGB maps for the NEUSA region with $R^2$ of 0.82 and MAE of 245.23 Mg $ha^{-1}$. Unfortunately, direct ecological interpretation of the GSE dataset remains challenging due to the black-box nature of the embedding-derived features [17]. Several AlphaEarth GSE features exhibited moderate relationships with LiDAR-derived canopy structural metrics according to the correlation analysis. Distinct embedding features such as A24, A36, A41, A49, and A61 consistently showed structured relationships with LiDAR variables associated with upper canopy height. This suggests that the embeddings captured underlying spatial and ecological patterns related to forest vertical structure beyond direct spectral information alone. Embeddings such as A36 and A49 repeatedly exhibited high loadings across several PCs. This is evident that these embedding features represented multiple dimensions of environmental and forest structural variability.

However, LiDAR-only AGB models outperformed GSE-only counterparts achieving 17.91% higher $R^2$ and more than 20.16% lower RMSE and MAE under limited sample size conditions. Both RF1_T and XGT_3 reached $R^2$ values closer to 0.79, highlighting the importance of including LiDAR structural variables in AGB estimation. Moran's I values remained below 0.0809 in all combined models and approached zero in the XGBoost model (XGT_3) after incorporating spatial predictors ($I = -0.0095$, $p > 0.05$). This indicates that combining canopy structural metrics with GSE representations captured both local forest structure and broader regional ecological gradients [44]. Therefore, we recommend integrating LiDAR metrics with GSE-derived features whenever both datasets are available for the smaller sample sizes. Because their integration consistently produced the strongest AGB estimation performance when the sample size is less than 600. Our results further demonstrate the potential of GSE data as a scalable alternative for regional AGB estimation in areas with limited LiDAR coverage, particularly when annual forest inventory data are available.

Also, we propose that future studies should further incorporate climatic, topographic, and environmental variables, such as temperature, precipitation, elevation, and slope to better represent ecological gradients influencing forest productivity and biomass variability [5, 58, 59]. Furthermore, we emphasize the importance of spatially representative regional inventory frameworks for successful geospatial AGB modeling. According to Min et al. [21] AGB related citation counts decreased more than 85% from 2015 to 2024. These declining trends may partially reflect the limited availability of regional datasets comparable to FIA plot coverage. Although the NEFIN CFI dataset provided a valuable foundation for regional biomass modeling, we recommend broader and more generalized coverage for high resolution geospatial studies.

Our study represents an important first step toward developing annually updated and publicly accessible regional biomass maps without revealing sensitive training plot locations. Such approaches are critically important for supporting transparent forest carbon accounting, climate mitigation planning, ecosystem monitoring, and large area ecological assessments [58, 55]. Publicly available biomass products can substantially improve accessibility for researchers, land managers, and policy makers, enabling consistent monitoring of forest carbon dynamics.

## Conclusion

Our study is one of the first efforts to systematically investigate the utilization of AlphaEarth's Google satellite embedding dataset, aerial LiDAR, and NEFIN continuous forest inventory data. Our findings demonstrate that GSE data can serve as an alternative data source for regional AGB estimations where LiDAR coverage is limited with $R^2$ values reaching up to 0.82. Combining LiDAR structural metrics with GSE variables improved the regional AGB estimation across northeastern US forests for small sample sizes ($n<600$). Among all evaluated configurations where $n <600$, the tuned XGBoost model (XG3_T) produced the strongest overall performance ($R^2 = 0.79$, MAE = 227.81 Mg $ha^{-1}$, RBias = 10.72%). which highlights the complementary value of structural canopy information with GSE data for small sample sizes. The GSE-only models outperformed LiDAR-only models when utilizing the full available GSE time series, showing an approximately 4% increase in $R^2$ and 9% of reductions in RMSE and MAE values. We conclude that LiDAR and GSE combined models are most suitable for applications requiring high-accuracy local or regional biomass estimation for smaller sample sizes. However, the strong performance of the GSE-only models ($R^2 = 0.82$) demonstrates the potential of embedding-derived features as an alternative dataset in regions where LiDAR coverage is limited. The scalability and global consistency of satellite embedding products further support their applicability for large area AGB monitoring and carbon assessment. In our study, several plot locations contained unusually high biomass values, which contributed to increased prediction uncertainty. Therefore, this study emphasizes the necessity of geospatially, ecologically and silviculturally representative forest inventory frameworks for reliable AGB estimation when using geospatial data. As vegetation remote sensing increasingly advances toward AI-based carbon monitoring applications, accurate AGB estimation depends on high quality inventory datasets as much as remote sensing data.

**Acknowledgements**

S.L. and C.W. acknowledge support from the McIntire-Stennis Capacity Grant (2025-2026), which funded this project. The authors are grateful to the Northeastern Forest Inventory Network (NEFIN) and its partnering inventory programs for providing access to forest inventory data. Special thanks are extended to Soren Donisvitch and the Forest Ecosystem Monitoring Cooperative at the University of Vermont, Shane Miller of Baxter State Park, and Lisa St. Hilaire of the Maine Natural Areas Program and The Nature Conservancy in Maine for their assistance in coordinating, facilitating, and releasing inventory datasets used in this study.

**Funding**

This work was supported by the USDA McIntire–Stennis Capacity Grant (2025–2026).

## References

1. Pan, Y., Birdsey, R.A., Fang, J., Houghton, R., Kauppi, P.E., Kurz, W.A., Phillips, O.L., Shvidenko, A., Lewis, S.L., Canadell, J.G. and Ciais, P., 2011. A large and persistent carbon sink in the world's forests. science, 333(6045), pp.988-993.
2. Hoover, Katie, and Anne A. Riddle. 2022. *U.S. Forest Carbon Data: In Brief*. Version 10. Congressional Research Service, Report No. R46313. Accessed November 5, 2025. https://crsreports.congress.gov.
3. Alongi, D.M., 2012. Carbon sequestration in mangrove forests. *Carbon management*, *3*(3), pp.313-322.
4. Catanzaro, P. and D'Amato, A.W., 2019. *Forest carbon: an essential natural solution for climate change*. University of Massachusetts Amherst.
5. Lamahewage, S.H.G., Witharana, C., Riemann, R. *et al.* Aboveground biomass estimation using multimodal remote sensing observations and machine learning in mixed temperate forest. *Sci Rep* 15, 31120 (2025). https://doi.org/10.1038/s41598-025-15585-6
6. Hudak, Andrew T., Patrick A. Fekety, Van R. Kane, Robert E. Kennedy, Steven K. Filippelli, Michael J. Falkowski, Wade T. Tinkham et al. "A carbon monitoring system for mapping regional, annual aboveground biomass across the northwestern USA. "Environmental Research Letters15, no. 9 (2020): 095003.
7. Sheridan, R.D., Popescu, S.C., Gatziolis, D., Morgan, C.L. and Ku, N.W., 2014. Modeling forest aboveground biomass and volume using airborne LiDAR metrics and forest inventory and analysis data in the Pacific Northwest. *Remote Sensing*, *7*(1), pp.229-255. https://doi.org/10.3390/rs70100229
8. Chojnacky, D.C., Heath, L.S. and Jenkins, J.C., 2014. Updated generalized biomass equations for North American tree species. *Forestry*, *87*(1), pp.129-151.
9. Jenkins, J.C., Chojnacky, D.C., Heath, L.S. and Birdsey, R.A., 2003. National-scale biomass estimators for United States tree species. *Forest science*, *49*(1), pp.12-35.
10. Tamiminia, H., Salehi, B., Mahdianpari, M., Beier, C. M., Johnson, L., and Phoenix, D. B.: A COMPARISON OF DECISION TREE-BASED MODELS FOR FOREST ABOVE-GROUND BIOMASS ESTIMATION USING A COMBINATION OF AIRBORNE LIDAR AND LANDSAT DATA, ISPRS Ann. Photogramm. Remote Sens. Spatial Inf. Sci., V-3-2021, 235–241, https://doi.org/10.5194/isprs-annals-V-3-2021-235-2021, 2021.
11. Johnson, L.K., Mahoney, M.J., Bevilacqua, E., Stehman, S.V., Domke, G.M. and Beier, C.M., 2022. Fine-resolution landscape-scale biomass mapping using a spatiotemporal patchwork of LiDAR coverages. *International Journal of Applied Earth Observation and Geoinformation*, *114*, p.103059.
12. Boudreau, J., Nelson, R.F., Margolis, H.A., Beaudoin, A., Guindon, L. and Kimes, D.S., 2008. Regional aboveground forest biomass using airborne and spaceborne LiDAR in Québec. *Remote Sensing of Environment*, *112*(10), pp.3876-3890.
13. Puliti, S., Breidenbach, J., Schumacher, J., Hauglin, M., Klingenberg, T.F. and Astrup, R., 2021. Above-ground biomass change estimation using national forest inventory data with Sentinel-2 and Landsat. *Remote sensing of environment*, *265*, p.112644.
14. Xu, G., Wu, S., Xu, C., Yang, X., Du, Y., Wang, G., Long, J. and Lin, H., 2026. Mapping Forest Aboveground Carbon Storage by Integrating Multi-Source Optical and Multi-Temporal Sentinel-1 SAR Data in Mixed Broadleaf–Coniferous Forests. *Remote Sensing*, *18*(4), p.570.

15. Nandy, S., Srinet, R. and Padalia, H., 2021. Mapping forest height and aboveground biomass by integrating ICESat-2, Sentinel-1 and Sentinel-2 data using Random Forest algorithm in northwest Himalayan foothills of India. Geophysical Research Letters, 48(14), p.e2021GL093799. https://doi.org/10.1029/2021GL093799
16. Brown, C. F., Kazmierski, M. R., Pasquarella, V J., Rucklidge, W. J., Samsikova, M., Zhang, C., Shelhamer, E., Lahera, E., Wiles, O., Ilyushchenko, S., Gorelick, N., Zhang, L. L., Alj, S., Schechter, E., Askay, S., Guinan, O., Moore, R., Boukouvalas, A., & Kohli, P.(2025). AlphaEarth Foundations: An embedding field model for accurate and efficient global mapping from sparse label data. arXiv preprint arXiv.2507.22291. doi:10.48550/arXiv.2507.22291
17. Google DeepMind (2025) *AlphaEarth Foundations helps map our planet in unprecedented detail*. Available at: https://deepmind.google/blog/alphaearth-foundations-helps-map-our-planet-in-unprecedented-detail/?utm source= chatgpt.com (Accessed: 5 May 2026).
18. Google Earth Engine (2025) *Introduction to the Satellite Embedding Dataset*. Available athttps://developers.google.com/earth-engine/tutorials/community/satellite-embedding-01-introduction (Accessed: 3 April 2026).
19. Khan, H. and Ahmad, A., 2025. Evaluating AlphaEarth Foundation Embeddings for Pixel-and Object-Based Land Cover Classification in Google Earth Engine.
20. Jin, C., Jiang, X., Wen, L., Wu, C., Xu, X. and Jiao, J., 2026. Assessing the Utility of Satellite Embedding Features for Biomass Prediction in Subtropical Forests with Machine Learning. *Remote Sensing*, *18*(3), p.436..
21. Tamiminia, H., Salehi, B., Mahdianpari, M., Beier, C.M., Johnson, L. and Phoenix, D.B., 2021. A comparison of decision tree-based models for forest above-ground biomass estimation using a combination of airborne lidar and landsat data. *ISPRS annals of the photogrammetry, remote sensing and spatial information sciences*, *3*, pp.235-241.
22. Min, X., Yusof, M.J.M., Fan, L. and Maruthaveeran, S., 2026. Vegetation Carbon Stock Estimation Using Remote Sensing: A Bibliometric and Critical Review. *Forests*, *17*(4), p.503.
23. Lamahewage, S. H. G., Witharana, C., Riemann, R., Fahey, R., & Worthley, T. (2025). Comparing Machine Learning and Statistical Models for Remote Sensing-Based Forest Aboveground Biomass Estimations. *Forests*, *16*(9), 1430. https://doi.org/10.3390/f16091430
24. Tang, H., Ma, L., Lister, A. J., O'Neil-Dunne, J., Lu, J., Lamb, R., Dubayah, R. O., & Hurtt, G. C. (2021). *LiDAR Derived Biomass, Canopy Height, and Cover for New England Region, USA, 2015* (Version 1). ORNL Distributed Active Archive Center. https://doi.org/10.3334/ORNLDAAC/1854 Date Accessed: 2026-05-05
25. Srivastava, H. and Pant, T., 2023, July. Evaluation of ESA CCI Above Ground Biomass Using Multi-Source, Multi-Spectral Data And Machine Learning. In *IGARSS 2023-2023 IEEE International Geoscience and Remote Sensing Symposium* (pp. 3082-3085). IEEE.
26. Ma, L., G. C. Hurtt, H. Tang, R. Lamb, E. Campbell, RO Dubayah, M. Guy et al. "Forest Aboveground Biomass and Carbon Sequestration Potential, Northeastern USA." ORNL DAAC (2022).
27. Blackard, J.A., Finco, M.V., Helmer, E.H., Holden, G.R., Hoppus, M.L., Jacobs, D.M., Lister, A.J., Moisen, G.G., Nelson, M.D., Riemann, R. and Ruefenacht, B., 2008. Mapping US forest biomass using nationwide forest inventory data and moderate resolution information. *Remote sensing of Environment*, *112*(4), pp.1658-1677.

28. Menlove, J. and Healey, S., 2021. CMS: Forest Aboveground Biomass from FIA Plots across the Conterminous USA, 2009-2019. ORNL DAAC.
29. U.S. Census Bureau. (2021). *State area measurements and internal point coordinates*. Accessed June 12, 2025, from https://www.census.gov/geographies/reference-files/2010/geo/state-area.html
30. PRISM Group, Oregon State University (2026) *PRISM Climate Data*. Available at: https://prism.oregonstate.edu (Accessed: 7 May 2026).
31. Ducey, M., Johnson, K., Belair, E., & Mockrin, M., 2016. Forests in Flux: The Effects of Demographic Change on Forest Cover in New England and New York. . https://doi.org/10.34051/p/2020.261.
32. Tang, G., Beckage, B., & Smith, B., 2014. Potential future dynamics of carbon fluxes and pools in New England forests and their climatic sensitivities: A model-based study. Global Biogeochemical Cycles, 28, pp. 286 - 299. https://doi.org/10.1002/2013gb004656.
33. University of Vermont, 2024. *Northeastern Forest Inventory Network (NEFIN)*. [online] Available at: https://www.uvm.edu/femc/nefin/ [Accessed 7 Jul. 2025].
34. Chojnacky, D. C., and L. S. Heath. 2002. Estimating down dead woody material in the United States. General Technical Report NE-298. U.S. Department of Agriculture, Forest Service.
35. Johnson, Kristofer D., Richard Birdsey, Andrew O. Finley, Anu Swantaran, Ralph Dubayah, Craig Wayson, and Rachel Riemann. "Integrating forest inventory and analysis data into a LIDAR-based carbon monitoring system." Carbon Balance and Management 9 (2014): 1-11.
36. Chen, N., Tsendbazar, N.E., Suarez, D.R., Silva-Junior, C.H., Verbesselt, J. and Herold, M., 2024. Revealing the spatial variation in biomass uptake rates of Brazil's secondary forests. *ISPRS Journal of Photogrammetry and Remote Sensing*, *208*, pp.233-244.
37. National Oceanic and Atmospheric Administration (NOAA), Office for Coastal Management. (2025). *Canopy – High Resolution Land Cover, Phase 1 Initial Layers* [Data set]. Available from https://coastalimagery.blob.core.windows.net/ccap-landcover/CCAP_bulk_download/High_Resolution_Land_Cover/Phase_1_Initial_Layers/Canopy/index.html (Accessed 13 November 2025).
38. U.S. Geological Survey. (2023, July 27). *National Land Cover Database (NLCD) 2021: Conterminous U.S. land cover* [Image]. U.S. Department of the Interior. https://www.usgs.gov/media/images/national-land-cover-database-nlcd-2021-conterminous-us-land-cover
39. Saim, A.A. and Aly, M.H., 2025. Machine Learning and Multisensor Data Fusion for Forest above Ground Biomass Estimation in Arkansas. *Earth Systems and Environment*, pp.1-18.
40. U.S. Geological Survey (2026) *USGS Lidar Explorer*. Available at: https://apps.nationalmap.gov/lidar-explorer/#/ (Accessed: 2 May 2026).
41. Wu, Q., 2021. lidar: A Python package for delineating nested surface depressions from digital elevation data. *Journal of Open-Source Software*, *6*(59), p.2965.
42. Fang, J., Wu, M., Zhang, Z. and Luo, W., 2025. Leveraging AlphaEarth Foundations Embeddings for High-Accuracy County-Scale Corn and Soybean Yield Estimation. *Authorea Preprints*.

43. Fu, W.J., Jiang, P.K., Zhou, G.M. and Zhao, K.L., 2014. Using Moran's I and GIS to study the spatial pattern of forest litter carbon density in a subtropical region of southeastern China. *Biogeosciences*, *11*(8), pp.2401-2409.
44. Ploton, P., Mortier, F., Réjou-Méchain, M., Barbier, N., Picard, N., Rossi, V., Dormann, C., Cornu, G., Viennois, G., Bayol, N. and Lyapustin, A., 2020. Spatial validation reveals poor predictive performance of large-scale ecological mapping models. *Nature communications*, *11*(1), p.4540.
45. Hengl, T., Nussbaum, M., Wright, M.N., Heuvelink, G.B. and Gräler, B., 2018. Random forest as a generic framework for predictive modeling of spatial and spatio-temporal variables. *PeerJ*, *6*, p.e5518.
46. Roberts, D.R., Bahn, V., Ciuti, S., Boyce, M.S., Elith, J., Guillera-Arroita, G., Hauenstein, S., Lahoz-Monfort, J.J., Schröder, B., Thuiller, W. and Warton, D.I., 2017. Cross-validation strategies for data with temporal, spatial, hierarchical, or phylogenetic structure. *Ecography*, *40*(8), pp.913-929.
47. Liu, X., Kounadi, O. and Zurita-Milla, R., 2022. Incorporating spatial autocorrelation in machine learning models using spatial lag and eigenvector spatial filtering features. *ISPRS International Journal of Geo-Information*, *11*(4), p.242.
48. Tamiminia, H., Salehi, B., Mahdianpari, M. & Goulden, T. State-wide forest canopy height and aboveground biomass map for new York with 10 m resolution, integrating GEDI, Sentinel-1, and Sentinel-2 data. Ecol. Inf. 79, 102404 (2024).
49. Devasahayam, S. and Albijanic, B., 2024. Predicting hydrogen production from co-gasification of biomass and plastics using tree based machine learning algorithms. Renewable Energy, 222, p.119883.
50. Su, Y., Wu, Z., Zheng, X., Qiu, Y., Ma, Z., Ren, Y. and Bai, Y., 2025. Harmonizing remote sensing and ground data for forest aboveground biomass estimation. Ecological Informatics, 86, p.103002.
51. Chen, J., 2025. Improving Tropical Forest Biomass Predictions with Multi-Output Deep Learning Models for Above-and Belowground Estimates. Informatica, 49(37).
52. Holman, J.O. and Hacherl, A., 2023. Teaching Monte Carlo Simulation with Python. Journal of Statistics and Data Science Education, 31(1), pp.33-44.
53. Wang, X., Gai, Y. and Huang, H., 2025. Design based Global Sensitivity Analysis for Quantity-Permutation Models. Technometrics, pp.1-12.
54. Hastie, T., Tibshirani, R. and Friedman, J., 2009. The elements of statistical learning.
55. Santoro, M., Cartus, O., Carvalhais, N., Rozendaal, D., Avitabile, V., Araza, A., De Bruin, S., Herold, M., Quegan, S., Rodríguez-Veiga, P. and Balzter, H., 2021. The global forest above-ground biomass pool for 2010 estimated from high-resolution satellite observations. Earth System Science Data, 13(8), pp.3927-3950.
56. Vorster, A.G., Evangelista, P.H., Stovall, A.E. and Ex, S., 2020. Variability and uncertainty in forest biomass estimates from the tree to landscape scale: the role of allometric equations. *Carbon Balance and Management*, *15*(1), p.8.
57. Winsemius, S., Babcock, C., Kane, V.R., Bormann, K.J., Safford, H.D. and Jin, Y., 2024. Improved aboveground biomass estimation and regional assessment with aerial lidar in California’s subalpine forests. *Carbon Balance and Management*, *19*(1), p.41.

58. Duncanson, L. I., Niemann, K. O., & Wulder, M. A. (2010). Integration of GLAS and Landsat TM data for aboveground biomass estimation. *Canadian Journal of Remote Sensing*, *36*(2), 129–141. https://doi.org/10.5589/m10-037
59. Lu, D., Chen, Q., Wang, G., Liu, L., Li, G. and Moran, E., 2016. A survey of remote sensing-based aboveground biomass estimation methods in forest ecosystems. *International Journal of Digital Earth*, *9*(1), pp.63-105.
60. Torre-Tojal, L., Bastarrika, A., Boyano, A., Lopez-Guede, J.M. and Grana, M., 2022. Above-ground biomass estimation from LiDAR data using random forest algorithms. *Journal of Computational Science*, *58*, p.101517.
61. Chen, Q., Laurin, G. V., Battles, J. J. & Saah, D. Integration of airborne lidar and vegetation types derived from aerial photography for mapping aboveground live biomass. Remote Sens. Environ. 121, 108–117 (2012).